\documentclass[12pt]{article}

\usepackage{graphicx}
\usepackage{color}
\usepackage[dvips]{epsfig}
\usepackage{bbm}
\usepackage{comment}
\usepackage[affil-it]{authblk}

\newcommand{\ket}[1]{\left\vert{#1}\right\rangle}

\date{}
\title{Order 3 Symmetry in the Clifford Hierarchy}
\author[1]{Ingemar Bengtsson}
\author[1]{Kate Blanchfield}
\author[2,3]{Earl Campbell}
\author[4]{Mark Howard}
\affil[1]{Stockholms universitet, Fysikum, S-106 91 Stockholm, Sweden}
\affil[2]{Department of Physics and Astronomy, University of Sheffield, Sheffield, \newline S3 7RH, UK}
\affil[3]{Dahlem Center for Complex Quantum Systems, Freie Universit\"{a}t Berlin, Berlin, Germany}
\affil[4]{Institute for Quantum Computing and Department of Applied Mathematics, University of Waterloo, Waterloo, Ontario, Canada, N2L 3G1}

\begin{document}
  \maketitle

\begin{abstract} 
We investigate the action of the first three levels of the Clifford hierarchy on sets of mutually unbiased bases comprising the Ivanovic MUB and the Alltop MUBs. Vectors in the Alltop MUBs exhibit additional symmetries when the dimension is a prime number equal to 1 modulo 3 and thus the set of all Alltop vectors splits into three Clifford orbits. These vectors form configurations with so-called Zauner subspaces, eigenspaces of order 3 elements of the Clifford group highly relevant to the SIC problem. We identify Alltop vectors as the magic states that appear in the context of fault-tolerant universal quantum computing, wherein the appearance of distinct Clifford orbits implies a surprising inequivalence between some magic states.
\end{abstract}

\newpage

\section{Introduction}

The Weyl-Heisenberg group, also known as the qudit Pauli group, is deeply interwoven with the very foundations of quantum mechanics \cite{Schwinger}. The group of unitaries that map the Weyl-Heisenberg group to itself under conjugation is known as the Clifford group, and plays a major role in a theory of fault-tolerant quantum computation \cite{Gottesman1}. If the dimension of Hilbert space is a prime number, then a complete set of mutually unbiased bases (MUB) arises as an orbit of the Clifford group \cite{Ivanovic, Vatan}. The vectors in such a set are also known as stabilizer states because they are stabilized by (i.e. they are eigenvectors of) elements of the qudit Pauli group.  

If a quantum computer is restricted to performing Clifford operations on the stabilizer states it cannot outperform its classical counterparts \cite{GK}. One way to achieve universal quantum computation is to introduce certain ``magic'' states for the quantum computer to act on \cite{Bravyi, Knill, Anwar, Campbell, Howard, Campbell2}. Such states can be obtained by acting on the stabilizer states with elements of the third level of the Clifford hierarchy, which consists of those unitaries that map the Weyl-Heisenberg group into the Clifford group \cite{Gottesman}. When acting on the stabilizer states, a unitary at this third level turns the usual MUB into another MUB, which is itself an orbit of the Weyl-Heisenberg group with one extra basis appended \cite{Alltop, Kate}. These are the magic states that we will discuss, under the name of Alltop vectors. In the qubit case they are also known as ``$H$-type magic states'' \cite{Bravyi}.

The purpose of this paper is to point out the peculiar role played by Clifford group elements of order 3 in this context whenever the dimension of Hilbert space is a prime $p = 1$ modulo 3. In this case each Alltop vector is invariant under such an order 3 element. What is more, the set of all Alltop vectors forms a configuration \cite{Hilbert} together with the set of largest subspaces left invariant by these order 3 elements. For a reason that we will come to these subspaces are called Zauner subspaces, and the precise statement is that each Alltop vector belongs to $p$ Zauner subspaces, and each Zauner subspace contains $2(p-1)$ Alltop vectors. 

One consequence is that the set of these magic states splits into 3 distinct orbits under the Clifford group if the dimension is $p = 1$ modulo 3, while there is only one orbit if $p = 2$ modulo 3. 
This is of interest to the magic state model for fault-tolerant universal quantum computing.
However, here our primary interest is a curious parallel between magic states and symmetric, informationally complete (SIC) measurements.
The latter form what are arguably the most distinguished of all Weyl-Heisenberg orbits. In dimension 2, the vectors in a SIC are also known as ``$T$-type magic states'' \cite{Bravyi}. In a general dimension $N$, a SIC is a POVM consisting of $N^2$ unit vectors $|\psi_I\rangle$ obeying 

\begin{equation} 
|\langle \psi_I|\psi_J\rangle |^2 = \frac{1}{N+1} 
\end{equation}

\noindent whenever $I \neq J$ \cite{Zauner,Renes}. 
With one exception, all known SICs are group covariant with respect to the Weyl-Heisenberg group and in prime dimensions this is the only group that could do the job \cite{Huangjun}.
It is an outstanding problem to prove that the Weyl-Heisenberg group produces SICs for all dimensions. The available evidence suggests that it does, but no constructive procedure is known \cite{Appleby, Scott, Zhu}. 

Analytic examples of SICs are known in (at the moment) 23 different dimensions. For some utterly mysterious reason all known SIC vectors are left invariant by a Clifford group element of order 3, in agreement with a conjecture first made by Zauner \cite{Zauner}. Moreover it appears that every Zauner subspace contains at least one SIC vector \cite{Appleby, Scott}. 
We now have a different line of argument singling out these subspaces for attention, and we suggest that this hints at a deeper connection between MUBs and SICs. Indeed, one weak link is already known \cite{ADF, Marcus}.
In dimension $p = 3$ there is a very direct link, effectively discovered by Hesse in a different language \cite{Hesse}, and elaborated on since \cite{Dang}. We believe that we have strengthened the case for such a link in dimensions of the form $p = 1$ modulo 3, and consequently these dimensions may be the most promising ones for solving the SIC existence problem. 

We review some known facts---briefly, because they are well explained elsewhere---in sections 2 and 3. In section 4 we point out the special role that the order 3 symmetries play within the Clifford hierarchy when the dimension equals 1 modulo 3. 
In section 5 we explore the consequences. In particular we show that the Zauner subspaces and the magic vectors in the MUBs form configurations whenever the dimension $p = 1$ modulo 3. We will also show that in these dimensions the magic vectors form three distinct orbits under Clifford gates. There is only one orbit if $p = 2$ modulo 3. In section 6 we comment on the remarkable reality properties of the Alltop vectors. Section 7 describes the relationship of Alltop vectors to quantum computing. Section 8 gives a brief summary.

Unless otherwise stated we assume that the dimension of Hilbert space is a prime number $p > 3$, since this obviates the need for complicated caveats. As a matter of fact MUBs work somewhat differently in prime power dimensions, and they may well not exist in dimensions not equal to a power of a prime while SICs presumably do.

\

\section{The Clifford hierarchy}

In prime dimensions there is an essentially unique unitary representation of the finite Weyl-Heisenberg group, generated by clock and shift operators 

\begin{equation} 
X|r\rangle = |r+1\rangle \ , \hspace{6mm} Z|r\rangle = \omega^r|r\rangle \ , \hspace{8mm} \omega \equiv e^{\frac{2\pi i}{p}} \ , 
\end{equation}

\noindent where the kets are labelled by integers modulo $p$. It is convenient to describe this group using the vector ${\bf p} = (p_1,p_2)$ and the displacement operators \cite{Appleby}

\begin{equation} 
D_{\bf p} = \omega^{\frac{p_1p_2}{2}}X^{p_1}Z^{p_2} \ . 
\end{equation}

\noindent We let $1/\beta$ denote the multiplicative inverse modulo $p$ of the integer $\beta$. The Clifford group also includes a copy of the symplectic group $SL(2,\mathbbm{Z}_p)$, whose defining representation consists of two by two matrices with entries that are integers modulo $p$ and whose determinant equals unity. It is generated by the matrices 

\begin{equation} 
T = \left( \begin{array}{cc} 1 & 1 \\ 0 & 1 \end{array} \right) \ , 
\hspace{8mm} F = \left( \begin{array}{rr} 0 & -1 \\ 
1 & 0 \end{array} \right) \ . 
\label{TF} 
\end{equation} 

\noindent They will figure later on. With the representation of the Weyl-Heisenberg group already fixed, the unitary representation of the symplectic group is 

\begin{equation} 
G = \left( \begin{array}{cc} \alpha & \beta \\ \gamma & \delta \end{array} 
\right) \hspace{2mm} \rightarrow \hspace{2mm} \left\{ \begin{array}{lll}
U_G = \frac{e^{i\theta}}{\sqrt{N}}\sum_{r,s} 
\omega^{\frac{1}{2\beta}(\delta r^2 - 2rs + \alpha s^2)} 
|r\rangle \langle s| & \ & \beta \neq 0 \\
\\
U_G = \pm \sum_s\omega^{\frac{\alpha \gamma}{2}s^2}|\alpha s\rangle \langle s| & \ & 
\beta = 0 \ . \end{array} \right. 
\label{1} 
\end{equation}

\noindent These operators are known as symplectic unitaries. We ignore the overall phase factors except to note that a suitable choice of $\theta$ means the unitary $U_G$ is of the same order as $G$. In fact, they can be chosen so that the unitary representation is faithful \cite{Appleby09}. The full Clifford group contains products of Weyl-Heisenberg and symplectic unitaries.

With the above definitions one finds 

\begin{equation} 
D_{\bf p}D_{\bf q} = \omega^{q_1p_2-q_2p_1}D_{{\bf p} + {\bf q}} \ , 
\hspace{8mm} U_GD_{\bf p}U^{-1}_G = D_{G{\bf p}} \ . 
\end{equation}

\noindent A general element of the Clifford group can be written as 

\begin{equation} 
C = \omega^kD_{\bf p}U_G \ , \hspace{8mm} G \in SL(2,\mathbbm{Z}_p) 
\ . \end{equation}

In what follows we will be particularly interested in order 3 
and order $p$ elements of the Clifford group so we outline some useful facts here. 
There is a link between the trace of $G$ and the order of $G$ \cite{Appleby}. In prime dimensions, $G$ is of order 3 if and only if $\mbox{Tr}(G)=-1$. 
Similarly, $G$ is of order $p$ if $\mbox{Tr}(G)=2$, unless $G$ is the identity matrix. Recall we have fixed the phase in Eq.~(\ref{1}) so that $U_G$ has the same order as $G$.

Clifford unitaries of order 3 have degenerate spectra. There are $p^3(p+1)$ order 3 Clifford elements when $p=1 \bmod 3$ and $p^3(p-1)$ when $p=2 \bmod 3$ \cite{Appleby,Huangjun}. They are called Zauner unitaries, after Zauner who conjectured their relevance to the SIC problem. Later we will confirm the number of Zauner unitaries in the former case using a simple counting argument involving hyperbolic M\"obius transformations on the projective line with $p+1$ elements.
It is also useful to note that in dimensions $p=1 \mbox{ mod } 3$ there exists an $H\in SL(2,\mathbbm{Z}_p)$ such that $H G H^{-1}$ is diagonal for all $G$ of order 3 \cite{Appleby}.

Clifford unitaries of order $p$ sometimes have degenerate spectra. We are interested in those with non-degenerate spectra. The Clifford group contains exactly $p(p+1)(p-1)$ cyclic subgroups of such elements \cite{Kate}. They relate to the Alltop MUBs, introduced in the next section.

Beyond the Clifford group---where we must venture to perform universal quantum computation---one may add further operators from the Clifford hierarchy \cite{Gottesman}. The whole hierarchy can be defined recursively.  
A unitary $U$ belongs to the $k^{\mathrm{th}}$ level of the Clifford hierarchy, if it does not belong to a lower level and for all elements $D_{\bf p}$ of the Weyl-Heisenberg group, we have that $UD_{\bf p}U^{\dagger}$ is an element of the $(k-1)^{\mathrm{th}}$ level of the Clifford hierarchy.
The third level of this hierarchy consists of operators that take operators in the Weyl-Heisenberg group to operators in the Clifford group under conjugation.
Since the Weyl-Heisenberg operators are of order $p$ and have a non-degenerate spectrum the targets must be operators of order $p$ that cannot be written as Weyl-Heisenberg translates. The third level of the Clifford hierarchy is not a group in itself and includes all operators of the form

\begin{equation}
U = C_{1} M^x C_{2} \ ,
\end{equation}

\noindent where $C_{1}$, $C_{2}$ are any Clifford unitaries, $x \in \left\{ 1,2,\ldots,p-1 \right\}$ and $M$ is given by

\begin{equation} 
M = \sum_r \omega^{r^3}|r\rangle \langle r| \ . 
\end{equation}

\noindent The $M$ stands for ``magic'' \cite{Campbell, Howard}. One finds 

\begin{equation} 
MD_{\bf p}M^{-1} = \omega^{-\frac{p_1^3}{2}}D_{\bf q}U_G \ , \label{eqn:MagicConj}
\end{equation}

\noindent where 

\begin{equation} 
{\bf q} = \left( \begin{array}{c} p_1 \\ p_2+3p_1^2 \end{array} \right) 
\ , \hspace{8mm} G = \left( \begin{array}{cc} 1 & 0 \\ 6p_1 & 1 \end{array} 
\right) \ . 
\end{equation}

\noindent The Clifford operation in Eq.~(\ref{eqn:MagicConj}) is of order $p$, and its spectrum is non-degenerate. There are altogether $p(p-1)(p^2-1)$ such Clifford group elements and they lie in $p(p+1)(p-1)$ cyclic subgroups \cite{Kate}. This entire conjugacy class can be obtained by repeatedly conjugating $D_p$ with $M$ and with suitable symplectic group elements.

It is worth noting that the set of all diagonal unitaries up to the third level of the hierarchy does form a group, generated by a displacement operator, an order $p$ symplectic unitary, and $M$. So the group is 
$Z_p\times Z_p\times Z_p$. We assume that $p > 3$, but analogues exist 
also for $p = 3$, and for $p = 2$ where the analogue of $M$ is known as the pi-over-eight gate. Then the analogous abelian subgroups are 
$Z_3\times Z_9$ and $Z_8$, respectively \cite{Howard}.   

\

\section{Mutually unbiased bases}

We now introduce a complete set of $p+1$ mutually unbiased bases 
(MUB), including the computational basis. 
The vectors in this %set of bases 
MUB are collectively known as stabilizer states.
The computational basis will be denoted $|I^{(0)}_a\rangle$ where $a \in \{0, 1, \dots ,p-1\}$ labels the vectors and $I$ stands for Ivanovi\'c \cite{Ivanovic}. One gets the rest of the MUB by adding $p$ bases obtained by acting with the generators of the symplectic group, defined in Eq.~(\ref{TF}). Thus, with integers $z \in \{ 0, 1, \dots , p-1\}$,  

\begin{equation} 
|I_a^{(z)}\rangle = (U_{T})^z|I^{(0)}_a\rangle \ , 
\hspace{8mm} |I_a^{(\infty )}\rangle = U_{F}|I^{(0)}_a\rangle \ . \end{equation} 

\noindent One can check that each of these bases is an eigenbasis of a maximally abelian subgroup of the Weyl-Heisenberg group, which is why they are mutually unbiased \cite{Vatan}. The MUB, or equivalently their labelling set $z \in \{0, 1, \dots , p-1 , \infty\}$, forms a finite projective line on which the symplectic group acts via M\"obius transformations, 

\begin{equation} 
G = \left( \begin{array}{cc} \alpha & \beta \\ \gamma & \delta \end{array} 
\right) \ , \hspace{8mm} z \rightarrow \frac{\alpha z + \beta }{\gamma z + \delta} \ . 
\end{equation}

\noindent The Weyl-Heisenberg group acts as the identity on this projective line; it permutes vectors within a basis. These M\"obius transformations are in many respects analogous to projective transformations of the real projective line. In particular they come in three types: hyperbolic with two fixed points, parabolic with one, and elliptic with no fixed point. A hyperbolic M\"obius transformation comes from a matrix $G$ that can be diagonalized by means of conjugation \cite{Appleby}. Order 3 operators give hyperbolic M\"obius transformations if and only if $p = 3k+1$.

\begin{figure}[ht]
\centering
\includegraphics[width=0.6\textwidth]{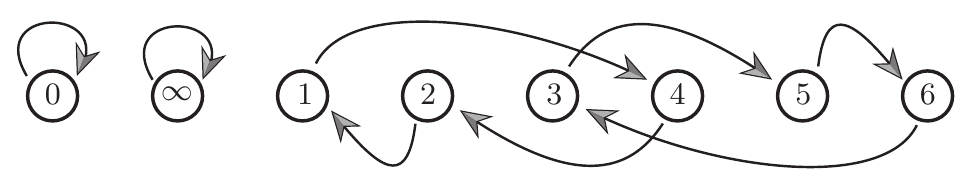}
\caption{\small{A M\"obius transformation of order 3 acting on a projective 
line with $7+1$ elements.}} 
\label{fig:mobius7}
\end{figure}

We can then count the number of order 3 symplectic unitaries by counting the possible ways of choosing pairs of fixed points. The first fixed point can be any basis in the MUB, for which we have $p+1$ choices. The second fixed point is a second basis, different to the first, for which we have $p$ choices. Avoiding double counting, this gives $p(p+1)/2$ possible fixed points. As the symplectic unitaries are order 3, $U_G^2$ has the same fixed points as $U_G$ and so overall we find $p(p+1)$ symplectic unitaries of order 3, as expected. Figure~\ref{fig:mobius7} shows a projective line for $p=7$ where each dot corresponds to a basis in the MUB.

Other MUBs are obtained by acting with operators from the third level of the Clifford hierarchy, such as 

\begin{equation} 
|A^{(z,x)}_a\rangle = M^x|I^{(z)}_a\rangle \ , \hspace{6mm} 
x \in \{ 1, \dots , p-1\} \ , \hspace{6mm} z \in \{0, 1, \dots , p-1 , \infty\} \ , 
\label{mubs} 
\end{equation}

\noindent where $A$ is for Alltop \cite{Alltop}. (The acronym MUBs is used for multiple complete sets of mutually unbiased bases.) Note that 

\begin{equation} 
|A_a^{(0,x)}\rangle \sim |I_a^{(0)}\rangle \ , \end{equation}

\noindent that is to say these two bases are equal up to irrelevant phases. Other Alltop MUBs are obtained by conjugating $M$ with elements of the Clifford group. To avoid burdening our notation too much we give only one example explicitly:

\begin{equation} 
|A^{\prime (z,x)}_a\rangle = U_FM^xU_F^{-1}|I_a^{(z)}\rangle \ , 
\hspace{8mm} 
|A^{\prime (\infty ,x)}_a\rangle \sim |I_a^{(\infty )}\rangle \ . 
\label{moremubs} 
\end{equation}

\begin{figure}[ht]
\centering
\includegraphics[width=0.7\textwidth]{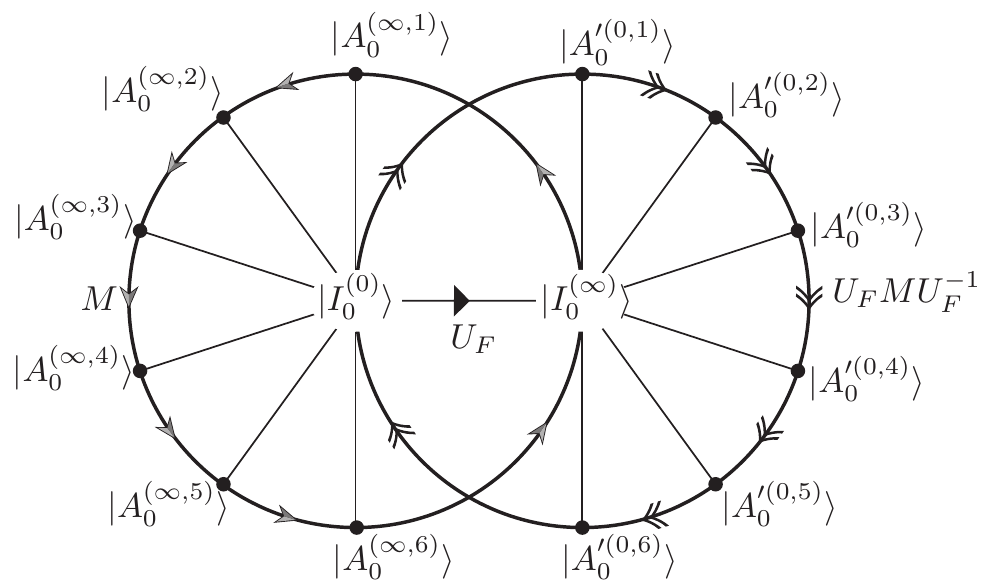}
\caption{\small{Alltop vectors created by the magical operator from the two fixed 
vectors in the Ivanovi\'c MUB, for $p = 7$.}} 
\label{fig:alltop}
\end{figure}

\noindent This particular Alltop MUB appears in section 6, where we show that some of the vectors in the MUB are real. Figure~\ref{fig:alltop} shows Alltop vectors generated by $M$ and $U_FM^xU_F^{-1}$ acting on two vectors in the Ivanovi\'{c} MUB.

Altogether this construction leads to $(p+1)(p-1)$ MUBs, each of which contains one basis from Ivanovi\'{c}'s MUB. The $p(p+1)(p-1)$ bases unique to the Alltop MUBs are eigenbases of cyclic subgroups containing order $p$ Clifford unitaries with non-degenerate spectra. We see that the number of these bases matches the number of such subgroups given in the previous section.
We refer to these $p^2(p+1)(p-1)$ vectors as Alltop vectors \cite{Kate}, although they are better known as ``magic states'' in the context of quantum computation \cite{Campbell, Howard}.

Eq.~(\ref{mubs}) reveals that the unitary equivalence between Ivanovi\'{c} and Alltop MUB constructions \cite{Godsil} is actually due to a unitary from the third level of the Clifford hierarchy. In addition, Eq.~(\ref{mubs}) will prove to be a key relationship for establishing that Ivanovi\'{c} and Alltop vectors lie in the same Zauner subspace $U_G$. Specifically, if $U_G|I^{(z)}_a\rangle=|I^{(z)}_a\rangle$ and $U_G$ and $M$ commute then

\begin{equation} 
U_G|A^{(z,x)}_a\rangle = U_GM^x|I^{(z)}_a\rangle   = M^xU_G|I^{(z)}_a\rangle =|A^{(z,x)}_a\rangle \ .
\end{equation}

\

\section{Order 3 symmetries within the Clifford hierarchy}

In the previous section we ended by highlighting the importance of symplectic unitaries $U_G$ that commute with the magic unitary $M$. A simple calculation verifies that

\begin{equation} 
MU_G = U_GM \hspace{5mm} \Leftrightarrow \hspace{5mm} 
G = \left( \begin{array}{cc} 
\alpha & 0 \\ \gamma & \alpha^2 \end{array} \right) \hspace{5mm} 
\mbox{where} \hspace{5mm} \alpha^3 = 1 \ \mbox{mod} \ p \ . 
\label{eqn:SuitableU} 
\end{equation}

\noindent To enumerate the number of commuting operators we must therefore 
find the number of integer solutions to the equation $\alpha^3 = 1$ 
modulo $p$. The solution $\alpha = 1$ always exists, and the resulting 
operators are of order $p$. It is a well known number theoretical fact that two additional solutions exist if and only if $p = 1$ modulo 3. The 
corresponding group elements are of order 3. Therefore order 3 symmetries 
have a special status in the hierarchy if and only if $p = 3k+1$ for some $k$.

One way of seeing why this case is singled out is to ask for cubic residues: integers $x$ of the form $x = y^3$ modulo $p$ for some integer $y$. If the dimension is $p = 3k+2$, then the set of cubic residues equals the whole field $\mathbbm{Z}_{p}$. We can see this by setting $y=x^{2k+1}$, which gives

\begin{equation}
y^3 = \left( x^{2k+1} \right)^3 = x^{6k+3} = x^{3k+2} x^{3k+1} = x^{p} x^{p-1} = x .
\end{equation}

\noindent The last step uses Fermat's Little Theorem, which states that $x^{p-1} = 1$ modulo $p$ whenever $x$ and $p$ are relatively prime. So every integer $y$ has a distinct cubic residue $x=y^3$ and thus the solution to $\alpha^3 = 1$ is unique.

If the dimension is $p = 3k+1$ on the other hand, we see that 

\begin{equation} 
(x^3)^k = (x^3)^{\frac{p-1}{3}} = x^{p-1} = 1 \ . 
\end{equation}

\noindent In this case, the cubic residues form a subgroup of order $k$ of the multiplicative group of integers modulo $p$, dividing it into 3 cosets. This fine structure will be important to us as we proceed. 

When $p = 3k +1$, each order 3 Zauner unitary has an eigenspace of dimension $k+1$, corresponding to the eigenvalue 1 once its overall phase has been suitably chosen \cite{Zauner}. It is believed that each such Zauner subspace contains a SIC vector (and it is known to be so for $p = 7,13,19$ \cite{Appleby, Grassl} and for $p = 31, 37, 43$ \cite{private}).

It is worth mentioning that a symplectic group element of the form in Eq.~(\ref{eqn:SuitableU}) is represented by a monomial unitary matrix---see Eq.~(\ref{1}). This simplifies the description of the corresponding Zauner subspaces, and it also simplifies the search for SIC vectors (in these dimensions) \cite{Appleby}.

\

\section{Order 3 symmetry of Alltop vectors}

We will now show that every Alltop vector is invariant under an order 3 element of the Clifford group, provided the dimension is $p = 3k+1$. We will focus on a particular such Zauner unitary, because all Zauner unitaries can be obtained from it by conjugating with symplectic unitaries and performing Weyl-Heisenberg translates \cite{Flammia}. We already know that in these dimensions any order 3 symplectic unitary lies in the same conjugacy class as a unitary that corresponds to a diagonal element of $SL(2,\mathbbm{Z}_p)$. Without loss of generality we therefore choose as our representative Zauner unitary

\begin{equation} 
{\cal Z} = \left( \begin{array}{cc} \alpha & 0 \\ 0 & \alpha^2 \end{array} \right) \ , \hspace{8mm} \alpha^3 = 1 \ , \hspace{5mm} \alpha \neq 1 \ . 
\end{equation} 

\noindent On the finite projective line of MUB it has fixed points at $z = 0, \infty$. From Eq.~(\ref{TF}) we know that the symplectic element $F$ interchanges these fixed points. We check that 

\begin{equation} 
M^{-1}U_{\cal Z}M = U_{\cal Z} \ , \hspace{8mm} 
U_F^{-1}U_{\cal Z}U_F = U^2_{\cal Z} \ . 
\end{equation}

\noindent The first expression is a direct consequence of Eq.~(\ref{eqn:SuitableU}). Then we observe (again using the explicit representation) that 

\begin{equation} 
U_{\cal Z}|I^{(0)}_0\rangle = |I^{(0)}_0\rangle \ , 
\hspace{8mm} U_{\cal Z}|I^{(\infty)}_0\rangle = |I^{(\infty)}_0\rangle \ . 
\end{equation}

\noindent The first vector in these two bases, but no other standard MUB vector, is left invariant. Using Eq.~(\ref{mubs}) it immediately follows that 

\begin{equation} U_{\cal Z}|A^{(\infty ,x)}_0\rangle = |A^{(\infty ,x)}_0\rangle \ . 
\end{equation}

\noindent Similarly, Eq.~(\ref{moremubs}) leads, after a minor calculation, to 

\begin{equation} U_{\cal Z}|A^{\prime (0 ,x)}_0\rangle = |A^{\prime (0 ,x)}_0\rangle \ . 
\end{equation}

\noindent The conclusion is that the Zauner subspace contains 2 standard and $2(p-1)$ Alltop vectors. The fact that the Alltop vectors have this extra invariance certainly singles out order 3 operators (in dimensions $p = 3k+1$) for special attention.

In the other direction, each Ivanovi\'{c} MUB vector is invariant under exactly $p^2$ Zauner unitaries, since there are $p$ choices for the second fixed point on the projective line, which fixes the symplectic part of the unitary, and then $p$ choices for the Weyl-Heisenberg part of the unitary as each basis vector is invariant under a cyclic subgroup of order $p$.

Putting everything together, there are $p(p+1)$ Ivanovi\'{c} MUB vectors, each one of which belongs to $p^2$ Zauner subspaces, and $(p+1)p^3/2$ Zauner subspaces each containing 2 Ivanovi\'{c} MUB vectors. This means that the Ivanovi\'{c} MUB constitutes what is known as a configuration of vectors and $(k+1)$-dimensional subspaces.

In the language of projective geometry, a collection of $m$ points and $n$ lines forms a configuration if each line contains $\pi$ points and each point has $\gamma$ lines passing through it. This leads to the condition $m \gamma = n \pi$, fulfilled by all configurations \cite{Hilbert}. This is expressed by the notation
\begin{equation}
\left( m_{|\gamma} , n_{|\pi} \right)  .
\end{equation}
For our purposes, a point corresponds to a vector (ray) in Hilbert space and a line corresponds to a subspace. 
The collection of Ivanovi\'{c} MUB vectors (stabilizer states) and Zauner subspaces of dimension $k+1$ can then be written as the configuration
\begin{equation} 
\left( p(p+1)_{|p^2}, \frac{(p+1)p^3}{2}_{|2}\right) \ . 
\end{equation}

\noindent More interestingly, the Alltop vectors also form a configuration. We know that each Zauner subspace contains $2(p-1)$ Alltop vectors so now we want to count how many Zauner unitaries leave each Alltop vector invariant. An order 3 Clifford group element leaving an Alltop vector invariant must also leave a special basis in the Ivanovi\'{c} MUB invariant. There are $p$ choices for the symplectic part, but this time there is no freedom in the Weyl-Heisenberg part because no WH operators leave the Alltop vector invariant.

Thus each Alltop vector belongs to $p$ Zauner subspaces, and we obtain the configuration 

\begin{equation} 
\left( (p+1)(p-1)p^2_{|p}, \frac{(p+1)p^3}{2}_{|2(p-1)}\right) \ . 
\end{equation} 

\noindent Configurations and their realizations in finite dimensional Hilbert spaces have a distinguished history in mathematics \cite{Hilbert, Dolgachev}, but we are not aware that this particular one has been encountered before. 

For $p = 3$, very similar considerations underlie the classical Hesse configuration \cite{Hesse}, which consists of 12 Zauner subspaces (orthogonal to 12 MUB vectors) and 9 SIC vectors at their intersections \cite{Dang}. However, this is a very special case because operators of order 3 are also of order $p$. Hesse's construction can be generalized to higher prime dimensions in two ways, the present one, and in the direction of the phase point operators introduced by Wootters \cite{Segre, Wootters}. The firm connection to SICs is lost in both versions, but the present version does focus on the mysterious order 3 Zauner symmetry in the SIC problem. 

From counting arguments it is clear that the Alltop vectors form a single orbit under the Clifford group if $p = 3k+2$, but three distinct orbits if $p = 3k+1$, since in this case the number of transformations leaving a particular Alltop vector invariant goes up with a factor of three. To see this explicitly let us ask how we can go between two Alltop vectors belonging to different Weyl-Heisenberg orbits but sitting in the same Zauner subspace. For this purpose we must use a symplectic unitary leaving a particular Ivanovi\'{c} basis, say the one with $z = 0$, invariant. This means that the matrix $G$ must have a zero in the upper right hand corner. One then checks that 

\begin{equation} 
G = \left( \begin{array}{cc} \alpha & 0 \\ \gamma & \delta \end{array} 
\right) \hspace{5mm} \Rightarrow \hspace{5mm} U_GM^x = M^{\frac{x}{\alpha^3}}U_G \ , \hspace{8mm} M^xU_G = U_GM^{x\alpha^3} \ . 
\end{equation} 

\noindent It follows that 

\begin{equation} 
U_G|A^{(0,x)}_a\rangle = U_GM^x|I^{(0)}_a\rangle = 
M^{\frac{x}{\alpha^3}}|I^{(0)}_{a'}\rangle = |A^{(0,\frac{x}{\alpha^3})}_{a'}\rangle \ . 
\end{equation}

\noindent In this way we create three multiplets corresponding to the three cosets into which the group of non-zero integers modulo $p$ is divided by the group of cubic 
residues.  

\

\section{Further symmetries of Alltop vectors}

For all $p > 3$ the Alltop vectors have a further symmetry, implemented by an anti-unitary operator belonging to the extended Clifford group. The latter is obtained by adjoining the matrix 

\begin{equation} G_K = \left( \begin{array}{cr} 1 & 0 \\ 0 & -1 \end{array} \right) 
\end{equation}

\noindent to the symplectic group, so that the determinant of the two-by-two matrices are allowed to take the values $\pm 1$ \cite{Appleby}. The matrix $G_K$ is represented in Hilbert space by means of complex conjugation, which we denote by $K$. An arbitrary anti-unitary operator can be written as $UK$, where $U$ is unitary and $K$ denotes complex conjugation in a given basis \cite{Wigner1}. Vectors invariant under an 
anti-unitary operator form a real subspace of Hilbert space---although it depends on the choice of basis whether the entries of these vectors are real numbers, or not. 
  
Consider also the unique order 2 element of the symplectic group, 

\begin{equation} 
A = F^2 = \left( \begin{array}{rr} - 1 & 0 \\ 0 & -1 \end{array} \right) \ 
\hspace{5mm} \Rightarrow \hspace{5mm} G_K A = F^{-1}G_K F \ .  
\end{equation}

\noindent The Weyl-Heisenberg translates of $A$ are identical to Wootters' phase point operators \cite{Wootters}, provided we choose the positive sign in Eq.~(\ref{1}), and we do so for convenience. The phase for the unitary matrix $U_F$ is chosen so that it becomes identical to the usual Fourier matrix. Each vector in the Ivanovi\'c MUB is invariant under an element of the Clifford group of order two, in particular the vectors $|I_0^{(z)}\rangle $ are invariant under $U_A$. The vectors $|I_0^{(0)}\rangle $ and $|I_0^{(\infty)}\rangle$ are special in that they are also invariant under complex conjugation. 

To see what this implies for the Alltop MUBs we again focus on the magic operator $M$. It commutes with the anti-unitary operator $U_AK$, 

\begin{equation} 
MU_{A}K = U_{A}KM \ . 
\end{equation}

\noindent For the Alltop MUBs that include the computational basis, it follows that 

\begin{equation} 
U_{A}K|A_0^{(\infty ,x)}\rangle = U_AKM|I_0^{(\infty )}\rangle = 
MU_AK|I_0^{(\infty )}\rangle = |A_0^{(\infty ,x)}\rangle \ . 
\end{equation}

\noindent There is another Alltop MUB obtained by conjugating the magic operator with $U_F$, namely 

\begin{equation} 
|A^{\prime (z,x)}_a\rangle = U_FM^xU_F^{-1}|I_a^{(z)}\rangle \ . 
\end{equation}

\noindent This MUB includes the Fourier basis and was given earlier in Eq.~(\ref{moremubs}). Using $U_A = U_F^2$ and $KU_F = U_F^3K$ it is easy to show that 

\begin{equation} 
K|A_0^{\prime (0 , x)}\rangle = |A_0^{\prime (0 , x)}\rangle  \ . \end{equation}

\noindent These particular Alltop vectors are manifestly real. The conclusion so far is that Alltop vectors lie in real subspaces. Each such real subspace contains $p-1$ distinct Alltop vectors, labelled by $x$.

If the dimension is $p = 3k+1$ we know that each Zauner subspace contains $2(p-1)$ Alltop vectors. Let us again focus on the representative Zauner operator $U_{\cal Z}$, which commutes with $M$, $K$ and $U_AK$. In this case, the $p-1$ Alltop vectors $|A_0^{(\infty ,x)}\rangle $ lie in the intersection of the Zauner subspace with the real subspace invariant under $U_AK$, and the $p-1$ Alltop vectors $|A_0^{\prime (0 ,x)}\rangle $ in its intersection with the manifestly real subspace. 

Thus the conclusion, when the dimension is a prime equal to 1 modulo 3, is that the $2(p-1)$ Alltop vectors in a given Zauner subspace are to be found in equal numbers in its intersections with two real subspaces. 
For $p = 7$, $19$ there also exist SIC vectors in these intersections \cite{Appleby}, but this does not seem to happen for any other value of $p$ \cite{Mahdad}.

\section{Relationship of Alltop vectors to Quantum Computation}

The set of Clifford unitaries and Pauli measurements are collectively known as stabilizer operations and these operations arise naturally in the context of fault-tolerant quantum computing (QC). The most promising proposals for building a large-scale quantum computer use error-correcting codes for which stabilizer operations are provably fault-tolerant (i.e., they do not introduce errors in an uncontrollable way) \cite{Gottesman1}. The stabilizer operations include all Clifford unitaries, Pauli measurements (with adaptive feedforward), preparation of stabilizer states and tracing out any subsystems. Unfortunately, any circuit using only stabilizer operations (applied to an initial computational-basis input state) is no more powerful than a classical computer, so some additional capability is needed. This fact has motivated the magic state model \cite{Bravyi,Knill} of QC whereby special resource states---magic states---are used up to implement non-Clifford unitaries. Stabilizer circuits supplemented by a supply of magic states are capable of universal and fault-tolerant quantum computing.

The connection to our work is via the Alltop vectors, which are precisely the magic states in prime dimensions. We have seen that the Alltop vectors are related to the Ivanovi\'{c} MUB by an element at the third level of the Clifford hierarchy. In quantum computing language, this statement becomes that the magic states are related to the stabilizer states by an element at the third level of the Clifford hierarchy. This was first studied for qubits \cite{Bravyi, Knill} and then extended to higher-dimensional qudit systems \cite{Campbell, Howard}. Furthermore, we have seen that up to Clifford unitaries these magic states split into either a single equivalence class (if $p=2 \bmod 3$) or three equivalence classes (if $p=1 \bmod 3$).

Clifford inequivalent quantum states vary in their potential as a computational resource in the magic states model, and this can be quantified by magic monotones. Veitch \textit{et. al.}~\cite{Veitch} proposed two such measures and here we comment on the relevance of one of them---the \textit{mana} $\mathcal{M}$---to our results.  The mana for an $n$ qudit state $\rho$ is easily calculated from
\begin{equation}
\mathcal{M}(\rho) = \ln \left(   \sum_{r=1}^{d^2}|W_{\rho}(r)| \right)
\end{equation}
where $W_{\rho}(r)$ denotes the components of the Wigner function \cite{Wootters}. From this definition follow numerous intuitive properties including
\begin{enumerate}
 \item For all stabilizer states $\rho_{\mathrm{stab}}$, the mana vanishes $\mathcal{M}(\rho_{\mathrm{stab}})=0$; 
 \item Positivity, for all $\rho$, $\mathcal{M}(\rho)\geq 0$;
 \item Additivity: for all $\rho$, $\sigma$, it follows $\mathcal{M}(\rho \otimes \sigma)=\mathcal{M}(\rho) + \mathcal{M}(\sigma)$;
 \item Clifford invariance: for all Clifford $C$, it holds that  $\mathcal{M}(C \rho C^{\dagger}) = \mathcal{M}(\rho)$;
 \item Monotonicity: for all trace preserving quantum channels $\mathcal{E}$ composed from stabilizer operations, and for all $\rho$,  we have $\mathcal{M}[\mathcal{E}(\rho)] \leq \mathcal{M}(\rho)$.
\end{enumerate}
These properties make mana especially relevant to quantum computation. Envisage a quantum circuit producing a multi-qudit state $\sigma$, and we must simulate this using only stabilizer operations and a supply of a magic states $\rho$. Combining monotonicity with additivity, we infer that one needs at least $n= \mathcal{M}(\sigma) / \mathcal{M}(\rho)$ copies of $\rho$.  The less mana contained in $\rho$, the more copies we need for a given computation. This naturally prompts the question ``how much mana do the Alltop vectors have?"  By Clifford invariance we know that all vectors $\ket{A^{(z,x)}_{a}}$ within the same Clifford orbit will have the same mana.  However, when $p=1 \bmod{3}$, it becomes possible for different Alltop vectors to carry more or less mana. For instance, when $p=7$ we can use $x=1,2,3$ as representatives of the three distinct equivalence classes of Alltop vectors, and find
\begin{eqnarray}
    \mathcal{M} \left( \ket{A^{(z,1)}_{a}} \right) & = &  0.8148 ,\\
    \mathcal{M} \left( \ket{A^{(z,2)}_{a}} \right) & = &  0.8148 ,\\
    \mathcal{M} \left( \ket{A^{(z,3)}_{a}} \right) & = &  0.8962 .
\end{eqnarray}
This hints that the vectors with $x=3$ might be a more powerful resource for quantum computation. One must be cautious about jumping to this conclusion since mana is not the unique magic monotone with the aforementioned properties.

Since we are also interested in SICs, we comment on the mana of SIC vectors. There are two Clifford orbits of SICs in $p=7$ and we find
\begin{eqnarray}
    \mathcal{M} (\psi_{\mathrm{SIC}}^{a}) & = &  0.8354 ,\\
    \mathcal{M} (\psi_{\mathrm{SIC}}^{b}) & = &  0.8116 ,
\end{eqnarray}
where $\psi_{\mathrm{SIC}}^{a}$ corresponds to the fiducial $7a$ and $\psi_{\mathrm{SIC}}^{b}$ corresponds to $7b$ in \cite{Scott}. In $p=7$, the state with maximal mana has $\mathcal{M} = 0.9022$ \cite{Gelo}.

\section{Conclusions}

In the context of quantum computation, the most interesting aspect of our current work is that it identifies an unexpected additional structure for magic states in prime dimensions of the form $p=1 \bmod 3$---not all magic states are created equal. In dimensions of the form $p=2 \bmod 3$, every magic state can be converted to an equivalent magic state by means of a Clifford gate, meaning that all these magic states exhibit the same amount of robustness to noise \cite{Howard} or mana (quantified as a resource \cite{Veitch}). In dimensions $p=1 \bmod 3$, the partitioning of Alltop vectors into three distinct Clifford orbits means this no longer holds true, and magic states from a particular orbit can be preferable to states from the remaining orbits.

We also aimed to throw a glimmer of light on the SIC existence problem. If the dimension of Hilbert space is a prime number $p$ it seems reasonable to hope for a connection to mutually unbiased bases, which---in their Alltop guise---form another distinguished orbit under the Weyl-Heisenberg group. For $p = 3$ the connection is very firm. A hint that a connection exists for all $p$ is known \cite{ADF, Marcus}, and prime dimensions do seem to be worth special attention \cite{Huangjun}. We observed that the Zauner subspaces---in which the SIC vectors are expected to lie---and the Alltop vectors form a configuration whenever $p = 1 \bmod 3$. 
In this case at least the order 3 Zauner symmetry plays a special role for Alltop and SIC orbits alike. We believe that this considerably strengthens the case for a connection. The Zauner subspaces are no longer featureless.

\section*{Acknowledgements:}

We thank David Andersson for illustrative discussions. EC acknowledges funding from the EU (SIQS). MH acknowledges financial support from FQXi, CIFAR and the Government of Canada through NSERC. This research was supported by the U.S. Army Research Office through grant W911NF-14-1-0103.

\

\end{document}